\documentclass[twocolumn,9pt]{article}
\usepackage{balance}
\usepackage{etoolbox}
\usepackage[export]{adjustbox}

\usepackage{flushend}
\usepackage[section]{placeins}
\usepackage[bf, footnotesize]{caption}
\usepackage{graphics}
\usepackage{CJKutf8}

\usepackage{amssymb}
\usepackage{mathtools, cuted}
\usepackage{widetext}

\usepackage{lineno}

\usepackage{multicol}
\usepackage{booktabs}
\usepackage{amsmath}
\usepackage{widetext}
\usepackage{flushend}
\usepackage[explicit]{titlesec}
\titleformat{\section}
  {\normalfont}{\thesection}{1em}{\MakeUppercase{#1}}
  \titleformat{\subsection}
  {\it}{\thesection}{1em}{{#1}}
  
\begin{document}
\small
\begin{strip}

\title{Physical description of the blood flow from the   internal jugular vein to the  right atrium of the heart: new ultrasound application perspectives}
\date{}

%% Authors and addresses/affiliations

\author{Francesco Sisini\\
\begin{small}
Department of Physics and Earth Sciences, University of Ferrara, Via Saragat 1, 44122 Ferrara, Italy
\end{small}\\\\
\begin{tiny}
Published on 5 March 2016.
\end{tiny}
}

%\address[Affil2]{Vascular Diseases Center, University of Ferrara, Via Aldo Moro 8, 44124 Cona (FE), Italy}
% Replace capitalized text with the appropriate information (use standard capitalization rules for your text, not all capitals.

\begin{CJK}{UTF8}{min}\Large 拳必殺  notes series\end{CJK}

\begin{center}
\line(1,0){450}
\end{center}
\maketitle

%% Do not remove the page break here.
\pagebreak

%\linenumbers

Ikken hissatsu (\begin{CJK}{UTF8}{min}拳必殺 )\end{CJK} means something like \textit{to annihilate at one blow}.
This document is part of a series of   notes each one targeting a single goal. Each note has  to \textit{annihilate  at one blow}! 

\section*{Conditions of use}

 This is a self-published methodological note distributed under the Creative Commons Attribution License (http://creativecommons.org/licenses/by/4.0/), which permits unrestricted use, distribution, and reproduction in any medium, provided the original work is properly cited. The note  contains an original  reasoning of mine and the goal to share thoughts and methodologies, not results. Therefore before using the contents of these notes, everyone is invited to verify the accuracy of the assumptions and conclusions.
\end{strip}
\section*{Background}
\begin{figure}
\includegraphics[scale=0.35,left]{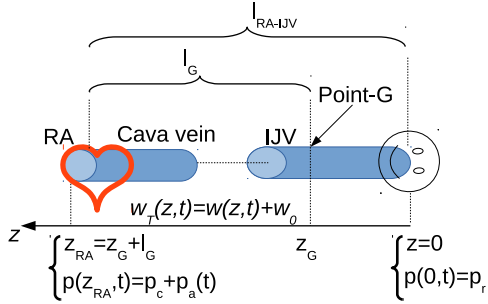} 
\caption{
%Modello geometrico del segmento cuore-giugulare. Il verso dell'asse \textit{z} \'e assunto positivo dalla giugulare verso il cuore. La distanza tra il cuore e il punto G dove si effettua la di misura ecografica \'e indicata con \textit{l}. La velocit\'a del sangue \'e indicata con $w$. 
Geometrical model of the jugular-heart segment. The \textit{z} axis is positive in the jugular-heart direction. \textit{l} is the distance between the RA and point G in which the scan takes place. Blood velocity is indicated by  $w$.
}
\label{fig:modello}
\end{figure}

The  brain drainage is due to the  venous blood flow  directed from the brain to the heart through the internal jugular veins (IJVs), epidural veins and vertebral veins (see Fig. \ref{fig:modello}). The brain-heart direction indicates that  there must be a negative pressure gradient  driving the blood flow (i.e. the pressure in the brain is higher than the pressure in the pressure in the right atrium (RA)).\\
The pressure, where the IJVs begin, is the residual arterial pressure and, despite its pulsatile nature, in this study it is considered to be constant.

The pressure in the RA varies  according to the cardiac cycle.  Its trace presents two wave peaks called  \textit{a} and  \textit{v} and two waves minima  called \textit{x} and  \textit{y}. Such  waves have a precise phase relationship with  the ECG waves PQRST \cite{Applefeld}, see Fig. \ref{fig:deltat}.
The waves \textit{a}, \textit{x}, \textit{v} and \textit{y} are also detectable at level of the neck due to the internal jugular vein (IJV) pulsation\cite{Mackenzie}.

% propagated to the internal jugular vein (IJV) and are detectable both as a pressure variation and as a variation of thir cross sectional area  (CSA)\cite{sisini}. 
%IJVs join with the cava vein and then with  the RA so it is natural to think that the change in pressure in the IJV is caused by the pressure change in RA. 

The pressure changes generated in the RA are indeed transmitted to the IJV and affect the velocity of the blood in two ways i) modifying the pressure gradient so that it is no longer constant over time and generates a non-steady flow \cite {sisini2016, Kalmanson}, ii ) cyclically varying the CSA of the IJV \cite{sisini}.\\ 
Since the walls of the IJVs are neither rigid nor collapsed, the pressure variations generated in the RA are transmitted to the IJV as a pressure wave with finite propagation velocity $ c $.
 Such pressure waves are responsible for the modulated component of the velocity of blood in the IJV. For this reason, their wave equation  can be used to derive the instantaneous velocity of the blood even in the absence of a direct measurement of such parameter obtained, for example,  by using an ultrasound  Doppler scanner. A first attempt to this goal was presented in \cite{sisinitoro}, where the instantaneous velocity of the blood in the IJV was determined qualitatively, for one cardiac cycle, using the Womersley equation   \cite{womera, womerb, womerc}.  

\subsection*{Physical description of the RA-IJV segment}
The RA-IJV system is represented in Fig. \ref{fig:modello} as a tube with a circular section.
G indicates a reference point on the right IJV. 
The  $w(z,r,t)$ function represents the velocity of the blood in the $z$ direction. It depends on $z$ coordinate, on $r$ (the distance from the axis $z$) and on $t$ (the time). The symbol $\overline{w}(z,t)$ is used to indicate the blood velocity averaged over the CSA.
The pressure at the  right end of the tube ($z=0$) is the residual arterial pressure and it is assumed to have a constant value $ p_r $ while  at the left end of the tube ($z=z_{RA}$) the pressure feels the effects of the RA activity   varying periodically with  the cardiac cycle. The pressure at $z=z_{RA}$  (see Fig. \ref{fig:modello}) is supposed to be the sum of a component $ p_c$  constant in time and of a periodic component $ p_a (t) $ which varies according to the atrial cardiac activity. The  blood flow is due to the pressure gradient between the ends  of the cylinder. It produces a    velocity $ \overline w_{T} (t)$ that results from  a constant component $  w_0$ due to the constant gradient $ \frac{p_c-p_r}{l_G} $    and a time varying component $ \overline w (t)$ due to the time varying $ \frac{p_a (t)-p_r}{l_G}  $.
 This system can be described using the Womersley equation that is a linear differential equation where the unknown  is the function $w$ and  the pressure gradient ($ \partial p (t) / \partial z) $ is the source term:

\begin{equation}
\label{eqmoto}
\frac{\partial^2 w(r,z,t)}{\partial r^2}+\frac{1}{ r}\frac{\partial w(r,z,t)}{\partial r}-\frac{\partial w(r,z,t)}{\partial t}=\frac{1}{\mu}\frac{\partial p(t,z)}{\partial z}
\end{equation}     
The radial component of the blood velocity and the non-linear terms of the equation are negligible for $c>>w$. \cite{womerc}
The solution of Eq.\ref{eqmoto} and the complete procedure to calculate it are explained in detail in\cite{sisinitoro}. However the solution is there defined up to the multiplicative constant $ c $ and up to the additive constant $w_0$. As a consequence, the instantaneous velocity trace, calculated in this way, is  proportional but not equal to the actual blood velocity.
\subsection*{Pressure waves propagation}
%La pressione si trasmette lungo la direzione AR-IJV secondo la seguente equazione
The pressure is transmitted along the direction AR-IJV according to the following equation
\begin{equation}
\label{waveequation}
\frac{\partial p}{\partial z}=+\frac{1}{c} \frac{\partial p}{\partial t}
\end{equation} 
For each cardiac cycle, the pressure in the RA reaches its maximum value  between $t_P$  and $t_Q$, where $t_P$ indicates the  ECG P wave and $t_Q$ indicates the  wave Q. In this note, for the sake of simplicity, it is assumed that the pressure in the RA is maximum in correspondence of the instant $t_{PQ}$ which is  the average  between $t_P$ and $t_Q$.
The relationship between the pressure in the RA and the pressure in the point G at the same instant of time is given by:
\begin{equation}
\label{sim}
p(t+l_{G}/c,z_G)=p(t,z_{RA})
\end{equation} 
Where  $z_G$ is the coordinate of the US scanning point and   $z_{RA}$ is the RA coordinate.

%The pressure in the RA   is maximum at  $t_{PQ}$, at the same instant, at the insonation point G of the IJV, the pressure $p(t_{PQ},z_G)$ will be equal to $p(t_{PQ}-l/c,z_{RA})$ as given by Eq. (\ref{sim}).

The pressure  is, at the same instant  $t_{PQ}$, maximum in the RA ($p(t_{PQ},z_{RA})$) , and equal to $p(t_{PQ}-l/c,z_{RA})$  at the insonation point G.
The qualitative JVP trace of pressure in the IJV , measured in G, can be obtained simultaneously with the acquisition of the ECG trace \cite{sisini2016}.  
On this   trace, that reports both JVP and ECG waves, the time interval $\Delta t_{aQP}$ between $t_{PQ}$ and $t_a$ is given by:
  
 \begin{equation}
 \label{gspot}
 \Delta t_{aQP}=t_a-t_{QP}=\frac{l_{G}}{c}
 \end{equation}
 where $t_a$ is the instant corresponding the \textit{a} wave.
 
%Il ritardo $\Delta t$ \'e  l'intervallo tra l'istante in cui si ha il massimo della pressione nel punto $G$ ($t_a$) e l'istante  in cui la pressione raggiunge il massimo nell'RA ($t_Q$). 
The  parameter $\Delta t_{aQP}$ is the  time interval   between the instant when the pressure is maximum at point G (instant $t_a$) and  the instant when the pressure is maximum at point RA (instant $t_{QP}$).

%La compliance $C$ della IJV \'e calcolabile usando l'equazione di Moens-Korteweg:
%\begin{equation}
%\label{MK}
%C=\frac{\rho}{A_0c^2}
%\end{equation}
%dove $A_0$ \'e la CSA della IJV in corrispondenza dell'onda \textit{x} del JVP. 

\begin{figure}
\includegraphics[scale=0.3]{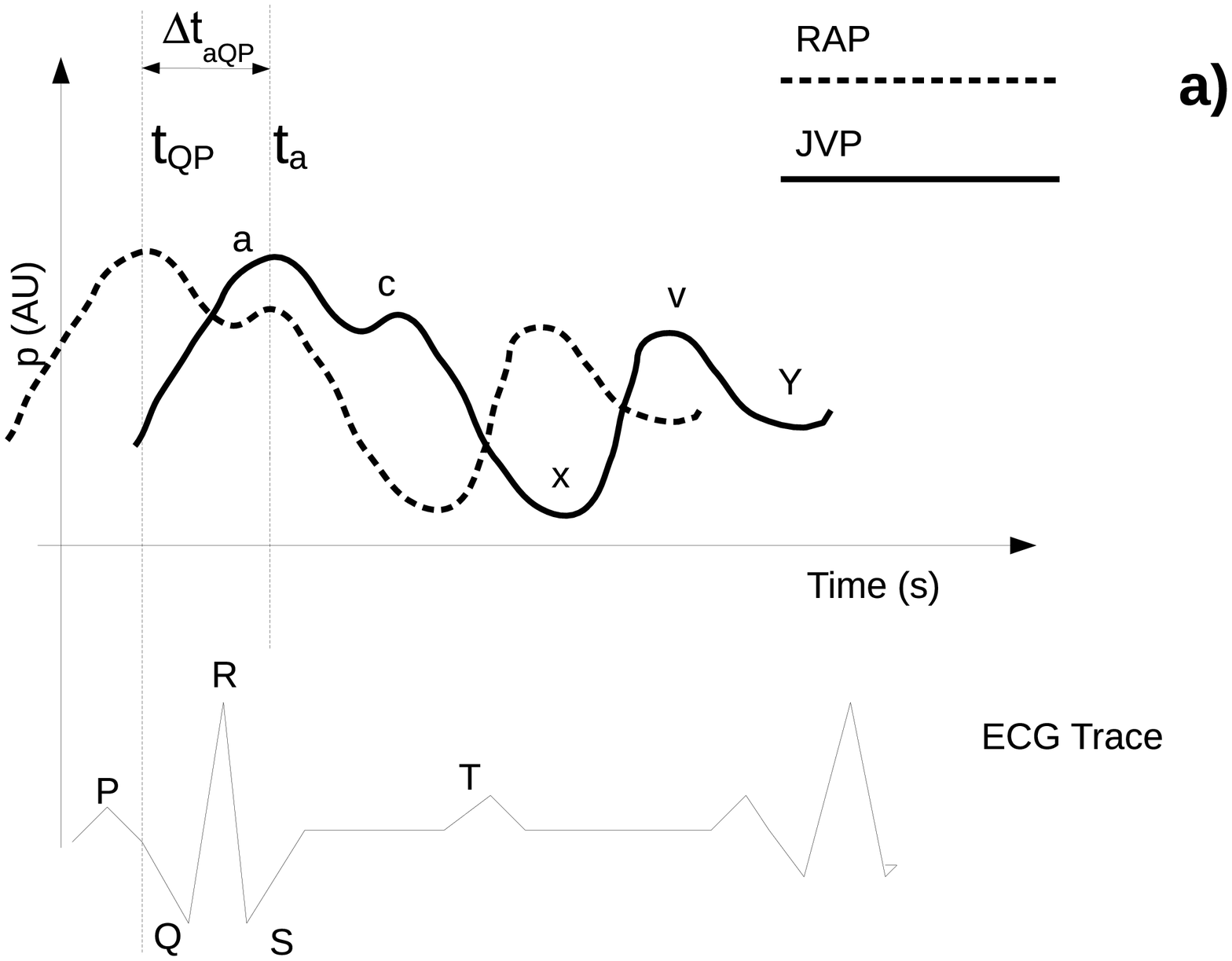} 
\includegraphics[scale=0.3]{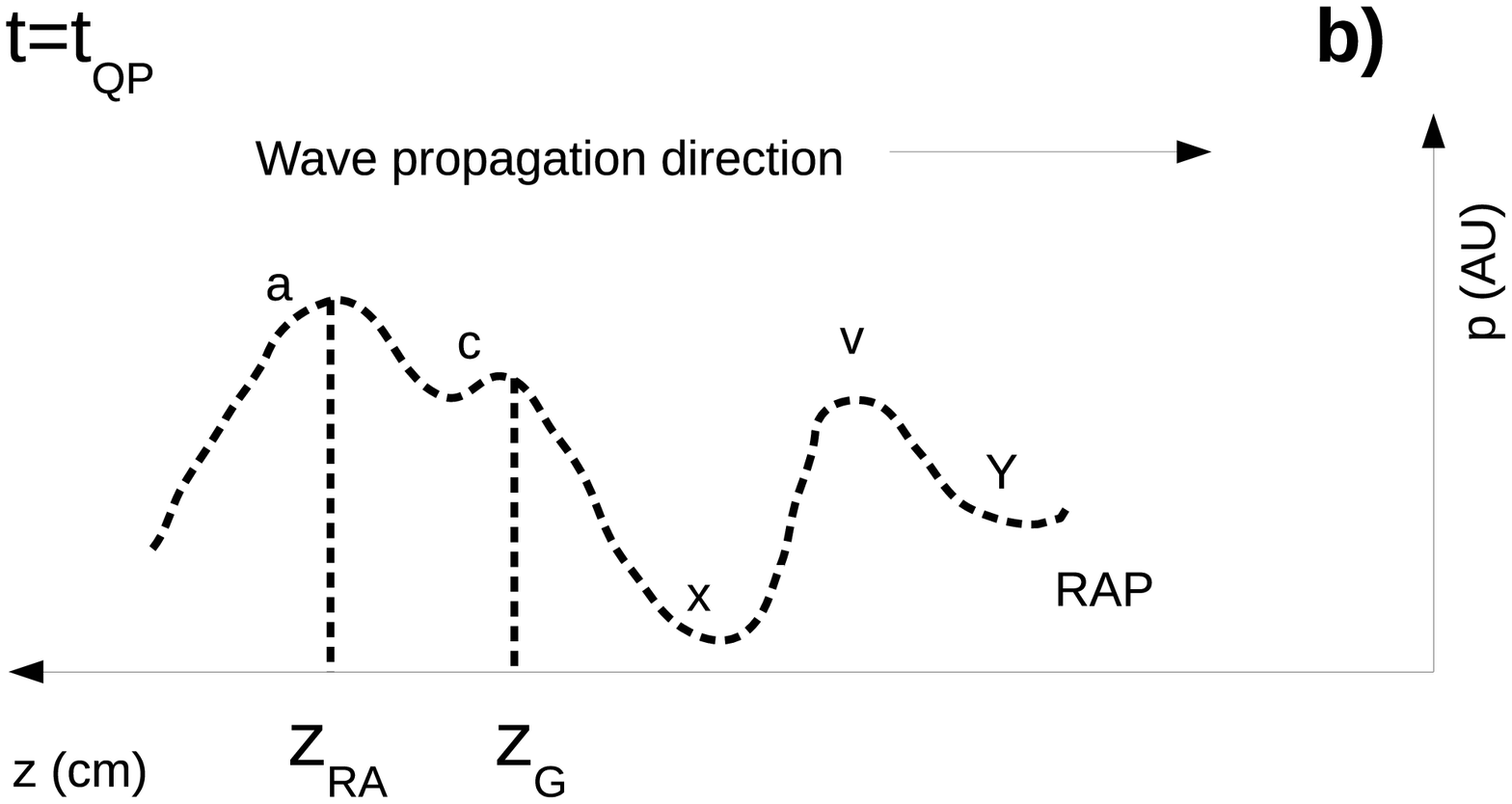} 
\caption{The figure \textbf{a)} shows the  JVP trace of pressure in the IJV , measured in G, together with the  ECG trace. The interval $\Delta t_{aQP}$ is represented over the curves \textit{a} and QP.  In figure \textbf{b)} it is shown the pressure along the \textit{z}-axes at instant $t_{QP}$.}

\label{fig:deltat}
\end{figure}

\section*{Methods}

 %l'applicabilit\'a della relazione   espressa in Eq. \ref{gspot}, \'e qui verificata confrontando la velocit\'a istantanea del sangue nella IJV calcolata secondo il modello visto sopra, con la velocit\'a istantanea misurata con uno scanner eco-Doppler.
The applicability of the relationship expressed in Eq. \ref{gspot}, can be tested by comparing the instantaneous velocity ($\overline{w}(t)$) of the blood in the IJV calculated according to the equation shown above, with the instantaneous blood velocity ($\overline{w}_D(t)$) measured with a Doppler scanner. However, the data and the results presented here are  for illustration only and are not to be considered the result of a scientific study.
 
%I valori della $CSA(t)$ e $\overline{w(t)}$ sono stati ottenuti rispettivamente dal  tracciato JVP+ECG  e dal tracciato Doppler di su un volontario sano e volontario	\cite{sisini2016}, vedi Fig. 
%The  $CSA(t)$ and $\overline{w}_D(t)$ dataset needed for the comparison are obtained by previously published results\cite{sisinitoro}. 

\subsection*{Pressure waves velocity calculation}
%Sul tracciato JVP+ECG sono stati identificati gli istanti $t_Q$ e  $t_a$ ed \'e stato calcolato l'intervallo $\Delta t$ tra i due. La misura $l$ \'e stata presa tra la quarta costola e il pomo di Adamo. La velocit\'a \'e stata calcolata come il rapporto tra  questi due parametri.
The delay $\Delta t_{aQP}$ is measured over the JVP+ECG trace as shown in Fig. \ref{fig:deltat}. The distance $l_G$  is measured on the volunteer's chest. The  velocity $c$ is the ratio between $l_{G}$ and $\Delta t_{aQP}$.

\subsection*{Compliance calculation}
From the Moens-Korteweg equation we obtain the  compliance per unit length:

\begin{equation}
\label{cp}
C^{'}=\frac{ CSA_{x} }{\rho c^2}
\end{equation}
where $CSA_{x}$ is the CSA of the IJV measured in G at the \textit{x} wave.
 
\subsection*{Pressure gradient calculation}
 %Il gradiente di pressione \'e stato calcolato moltiplicando il valore istantaneao della $CSA(t)$  per la compliance $C'$:

The pressure inside the IJV is calculated from the instantaneous $CSA$ as follows: 
\begin{equation}
\label{compliance} 
 p(t)=\frac{1}{C^{'}}CSA(t)
 \end{equation}
 substituting the expression of $p(t)$ obtained above into Eq. (\ref{waveequation}) we obtain the expression for the pressure gradient:
\begin{equation}
\label{dpz} 
  \frac{\partial p}{\partial z}=\frac{1}{cC^{'}}\frac{\partial CSA}{\partial t}
 \end{equation}

  \subsection*{Flow velocity calculation $\overline{w}(t,z)$}
%La velocit\'a longitudinale del flusso, mediata sulla sezione del vaso ($\overline{w}(t,z)$), \'e stata calcolata sostituendo al gradiente di pressione $\partial p/ \partial z$ in Eq.(\ref{eqmoto}) l'espressione trovata in Eq. (\ref{dpz}). I dettagli dell'algoritmo per la  soluzione sono riportati in\cite{sisinitoro}.
The function  $\overline{w}(t,z)$ is calculated by inserting into the  Eq.(\ref{eqmoto})
the expression for the pressure gradient found in  Eq. (\ref{dpz}). The mathematical details are given in\cite{sisinitoro}.
\begin{figure}
\includegraphics[scale=0.3]{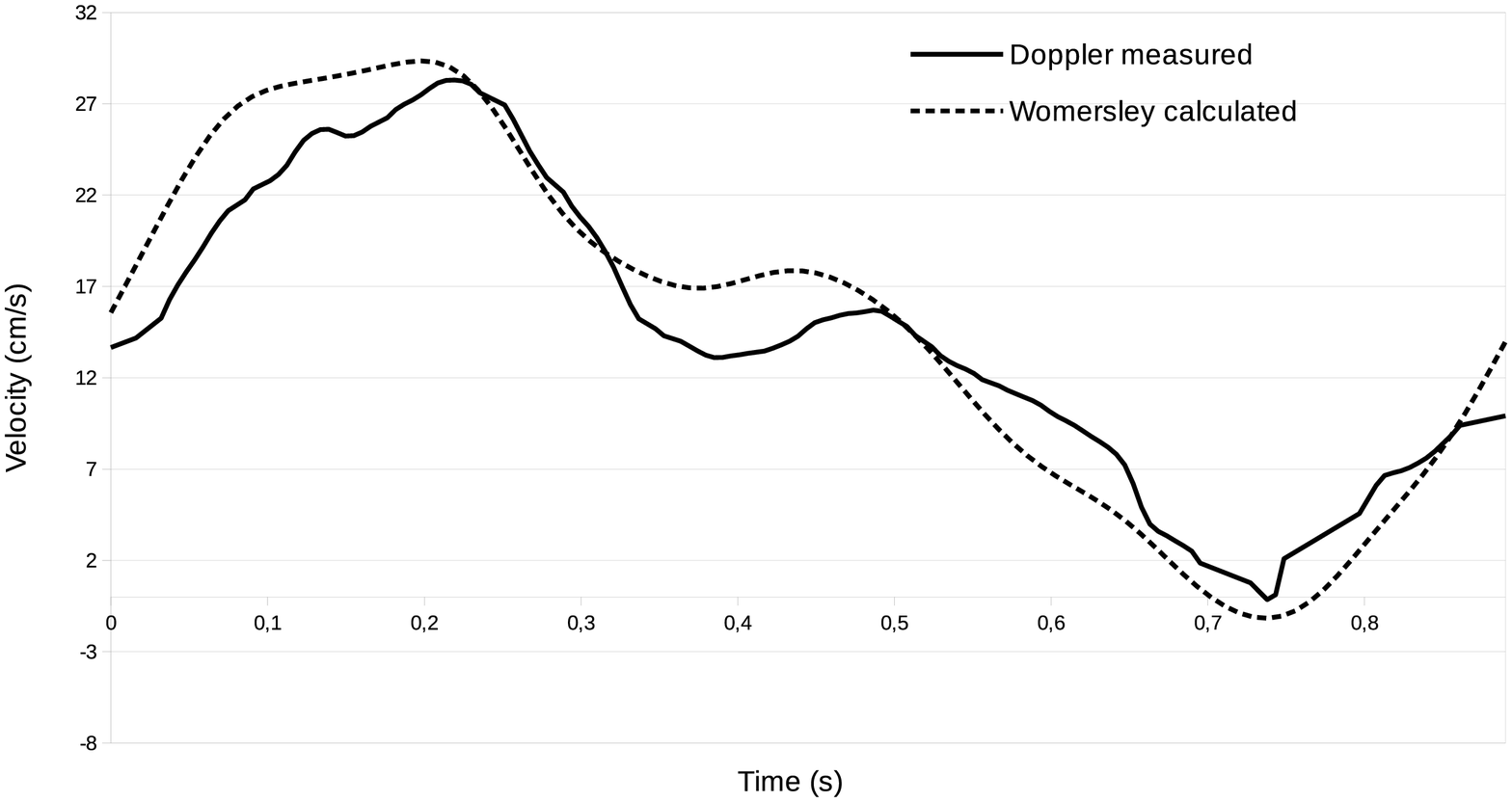} 
\caption{Calculated velocity trace  together with the experimental one obtained by the Doppler examination.}
\label{fig:velocity}
\end{figure}
\pagebreak
\section*{Results example}
\subsection*{Pressure waves velocity calculation}
%L'intervallo $\Delta t$ \'e risultato di 0.14 s, la distanza $l$ \'e risultata 23 cm. La velocit\'a $c$ \'e quindi 164 cm/s.
The measured delay ($\Delta t_{aQP}$) was 0.14 s and the distance $l_G$ was 23 cm. As a consequence the velocity $c$ was 164 cm/s.

\subsection*{Compliance calculation}
%Il valore minimo della CSA ($CSA_x$) \'e risultata 0.2 cm$^2$, la compliance per unit\'a i lunghezza \'e risultata pari a 9.8$\times 10^{-3}$ cm$^2/$mmHg
The minimum value of the CSA ($CSA_x$) during the cardiac cycle   was 0.2 cm$^2$ and the compliance for unit of length was  9.8$\times 10^{-3}$ cm$^2/$mmHg.

\subsection*{Flow velocity calculation $\overline{w}(t,z)$}
The velocity was calculated following Eq. (\ref{eqmoto}) and its trace is shown in Fig. \ref{fig:velocity} together with the experimental trace obtained by the Doppler examination. 
The two traces in agreement, nevertheless, since  the velocity  $\overline{w}(t,G+l)$ was calculated up to an additive constant,  only the wave amplitude has physical meaning whereas its shift along the velocity axis is not significant.   

\section*{Acknowledgement}
The author wants to thank Eleuterio Toro for his critical comments to the paper and Giacomo Gadda, Valentina Tavoni and Valentina Sisini to have kindly reviewed the manuscript.

%\section*{Discussion}
%The
%In questo lavoro  viene presentata la descrizione fisica del flusso sanguigno dalla IJV all'RA. Viene inoltre presentato un metodo inedito per calcolare la velocit\'a delle onde di pressione che dall'RA risalgono verso la IJV che si basa sull'analisi  simultanea dei tracciati JVP ed ECG. Tale metodo consente di calcolare la velocit\'a istantanea del sangue a meno di una costante additiva $w_0$, basandosi sul valore istantaneo della CSA misurato ad esempio con tecnica ultrasonografica.  

\pagebreak

\end{document}